\documentclass{aa}                        % default manuscript style (AA)
\usepackage{graphicx, latexsym}                   % Prefered graphics (AA)
\usepackage{natbib}                       % Bibliography package
\bibpunct{(}{)}{;}{a}{}{,}               % to follow the A&A style

\newcommand{\Mdot}{\hbox{$\dot M$}}
\newcommand{\Rsun}{\hbox{$R_\odot$}}
\newcommand{\Rstar}{\hbox{$R_*$}}
\newcommand{\Lsun}{\hbox{$L_\odot$}}

\newcommand{\Msunyr}{\hbox{$M_\odot\,$yr$^{-1}$}}
\newcommand{\Teff}{\hbox{$T_{eff}$}}

\newcommand{\Vinf}{\hbox{$V_\infty$}}

\newcommand{\kms}{\hbox{km$\,$s$^{-1}$}}

\newcommand{\ie}{\hbox{i.e.,}}

\newcommand{\etal}{\hbox{et~al.}}

\newcommand{\gnds}{\hbox{1s$^2$\,$^1$S}}
\newcommand{\tripp}{\hbox{1s\,2p\,$^3$P$^o$}}
\newcommand{\singp}{\hbox{1s\,2p\,$^1$P$^o$}}

\begin{document}

\title{On the sensitivity of \ion{He}{i}\ singlet lines to the \ion{Fe}{iv}\
model atom in O stars}

\author{F. Najarro \inst{1}
\and D. J. Hillier\inst{2}
\and J. Puls \inst{3}
\and T. Lanz \inst{4}
\and F. Martins \inst{5} }

\institute{ Instituto de Estructura de la Materia, Consejo Superior de 
Investigaciones Científicas, CSIC, Serrano 121, 28006 Madrid, Spain
\and
Dept. of Physics and Astronomy,
University of Pittsburgh, 3941 O'Hara St.,
Pittsburgh, PA 15260, USA
\and
Universit\"ats-Sternwarte M\"unchen, Scheinerstrasse 1, 81679
Munich, Germany
\and
Department of Astronomy, University of Maryland, College
Park, MD 20742, USA
\and
Max Planck Instit\"ut f\"ur Extraterrestrische Physik, 
Postfach 1312, 85741 Garching, Germany
}

\offprints{F. Najarro, \email{najarro@damir.iem.csic.es}}

\date{Received DD MMMM 200Y/ Accepted DD MMMM 200Y}

\abstract{}
{Recent calculations and analyses of
O star spectra have revealed discrepancies
between theory and observations, and between different
theoretical calculations, for the strength of optical
\ion{He}{i}\ singlet transitions.
We investigate the source of these discrepancies.}
{Using a non-LTE radiative transfer code we
have undertaken detailed test calculations for a range of O star
properties. Our principal test model has
parameters similar to those of the O9V star, 10 Lac.}
{We show that the discrepancies
 arise from uncertainties in the 
radiation field in the \gnds--\singp\ transition
near 584\AA. The radiation field at 584\AA\ is
influenced by model assumptions, such
as the treatment of line-blanketing and the
adopted turbulent velocity, and by the 
\ion{Fe}{iv} atomic data.  It is
shown that two \ion{Fe}{iv}\ transitions near
584\AA\ can have a substantial influence on the strength of the 
\ion{He}{i} singlet transitions.} 
{Because of the
difficulty of modeling the \ion{He}{i}\ singlet lines, particularly
in stars with solar metalicity, the \ion{He}{i}\
triplet lines should be preferred in spectral analyses.
These lines are much less sensitive to model assumptions.}

\titlerunning{\ion{He}{i} singlet lines in O stars}
\authorrunning{Najarro et al.}

\maketitle

\section{Introduction}
\label{section:intro}

With the advent of non-LTE line-blanketed model atmosphere calculations
(Hubeny and Lanz \citeyear{HL95_tlusty}; 
 Hillier \& Miller \citeyear{HM98_blank};
 Pauldrach \etal\  \citeyear{PHL01_WMBASIC}; 
 Gr\"afener \etal\ \citeyear{GKH02};
 Puls \etal\ \citeyear{PUS05_FW})
considerable advances in the analysis of massive stars have been made. 
There has, for example, been a significant downward revision in the 
effective temperature of O stars
(Martins et~al. \citeyear{MSH02_Teff}, \citeyear{MSH04_Teff_cal}; 
Crowther et~al. \citeyear{CHE02_FUSE};
Herrero et~al.\ \citeyear{HPN02_OB};
Bianchi \& Garcia \citeyear{BG02_Teff};
Garcia \& Bianchi \citeyear{GB04_Teff};
Repolust et~al.\ \citeyear{RPH04_Teff};
Massey \etal\ \citeyear{MBK04_LMC_O_stars}, \citeyear{MPP05_LMC_O_stars}). 
Moreover, the discrepancy between spectroscopic
and evolutionary masses (Groenewegen \etal\ \citeyear{GLP89_MD},
Herrero \etal\ \citeyear{HKV92_OBstars}) has been greatly reduced 
(e.g., Herrero \citeyear{Her03}; Repolust \citeyear{RPH04_Teff}).

A key diagnostic of the effective temperature in O stars is the
ratio of \ion{He}{i} line strengths relative to those of \ion{He}{ii}
(e.g., Kudritzki \etal\ \citeyear{KSH83_zpup}; Herrero \etal\ \citeyear{HPV00_CYGOB}).  
However detailed comparisons with observations
show discrepancies larger than expected. Further, theoretical
modeling reveals sensitivities of some lines to the adopted
techniques and assumptions while there are also discrepancies between 
results obtained using different codes. 

From very early on it was realized that in extended atmospheres
the \tripp\ and \singp\ states become overpopulated, and that
this affects the strength of the \ion{He}{i}\ lines
(e.g., Wellmann \citeyear{Wel52a_HeI}; \citeyear{Wel52b_HeI}; \citeyear{Wel52c_HeI};
Ghobros \citeyear{Gho62_HeI}). This is termed
the dilution effect, and was more recently discussed
by Voels \etal\ \cite{VBA89_dil_eff}.  With the advent of non-LTE 
line-blanketed model atmospheres it was expected that the new model
atmospheres would be capable of modeling the full \ion{He}{i}\
spectrum. However, as noted by Hillier \etal\ \cite{HLH03_AV83}, it was
not possible to get a simultaneous fit to all \ion{He}{i}
lines. Further, in that study of two O stars in the SMC, it was noted 
that the \ion{He}{i}\ singlet lines were more sensitive to 
model details than the \ion{He}{i}\ triplet line 4473\,\AA.
As additional studies have been undertaken it has become
clear that while model calculations often
have difficulty fitting the \ion{He}{i}\ singlet 
transitions they give satisfactory fits for the
\ion{He}{i}\ triplet lines (e.g., Puls et al. \citeyear{PUS05_FW}; 
Repolust et al. \citeyear{RPH05_IR}; Martins et~al. \citeyear{MSH05_WW};
Massey et~al. \citeyear{MPP05_LMC_O_stars}; 
Mokiem et~al. \citeyear{MKP05_GA}; Lanz \etal\ \citeyear{Lanz_10Lac}).

More recently, comparisons between CMFGEN (Hillier
and Miller \citeyear{HM98_blank}) and 
TLUSTY (Hubeny and Lanz \citeyear{HL95_tlusty}), and between
CMFGEN and FASTWIND (Puls \etal\ \citeyear{PUS05_FW}), 
have shown inconsistent predictions, in 
some parameter ranges, for the strength of the \ion{He}{i}
singlet lines (e.g., Puls \etal\ \citeyear{PUS05_FW}).  
The disagreement between CMFGEN and TLUSTY was somewhat surprising
since previous comparisons (e.g., Hillier \& Lanz \citeyear{HL01},
Bouret \etal\ \citeyear{BLH03_NGC346}) had shown excellent agreement.
However the comparisons were limited to a few models, and it is has
become apparent that the discrepancies only occur for
certain parameter ranges. Given the different assumptions in the codes,
different atomic data, and different techniques, it is not immediately obvious
which code is giving the correct answers. Agreement between observations and
CMFGEN was improved in some cases, for example, by adding additional
species (e.g., Argon, Neon, Nickel and Calcium), and by reducing
the turbulent velocity used in the atmospheric structure/population
calculations from 20 to 10\,\kms (Martins \etal\ \citeyear{MSH05_WW}).   

In this paper we discuss the analysis of \ion{He}{i}\ lines in O
stars, and the inconsistencies between triplet and singlet
lines, between observations and models, and between results 
obtained using different atmospheric codes.
The complicated behavior of the \ion{He}{i}\ singlet line strengths
is shown to be a consequence of complicated radiative transfer effects
in the neighborhood of the \ion{He}{i} 
\gnds\  -- \singp\ transition at 584\AA. In particular, it
is shown that two \ion{Fe}{iv}\ lines, which closely
coincide in wavelength with this transition
influence the population of the \singp\ level, and
thus the strength of \ion{He}{i}\ singlet lines that couple
to the \singp\ level. Indeed, a closer examination of the 
discrepancies reveals that the lines that are most discordant between
different model calculations are not the \ion{He}{i}\ singlet
lines per se, but rather the transitions involving \singp.

That line-line interactions, and the precise details of
line blanketing in the far UV, might affect optical
line diagnostics is not surprising. In planetary nebula
the fluorescence of \ion{O}{iii}\ lines by \ion{He}{ii}\ Ly$\alpha$
(the Bowen mechanism, Bowen \citeyear{Bow34})
is well documented (see, e.g., Osterbrock \citeyear{Ost89}).
In WN stars many  of the N lines are affected by continuum
fluorescence processes, and hence sensitive to blanketing
(Hillier \citeyear{Hil88_WN}). Further, 
Schmutz (\citeyear{Sch97_phot_loss})
suggested that photon loss from the \ion{He}{ii}\
resonance transition, by overlapping metal lines, could
significantly affect the ionization structure in W-R stars
and hence help to solve the problem of how to
accelerate W-R winds. The effect was investigated in 
the modeling the WC component of the WR+O binary $\gamma$ Velorum
(De Marco \citeyear{DSC02_GamVel}), and
found to be important. Martins \cite{Mar04_PhD} investigated the 
same mechanism in O star models (i.e., the influence
of metal lines near \ion{He}{ii}\ Ly$\alpha$) and also
found that the effects were significant.

\section{The interaction of \ion{He}{i} with \ion{Fe}{iv}}

\begin{figure}
\includegraphics[angle=270.0, scale=0.9]{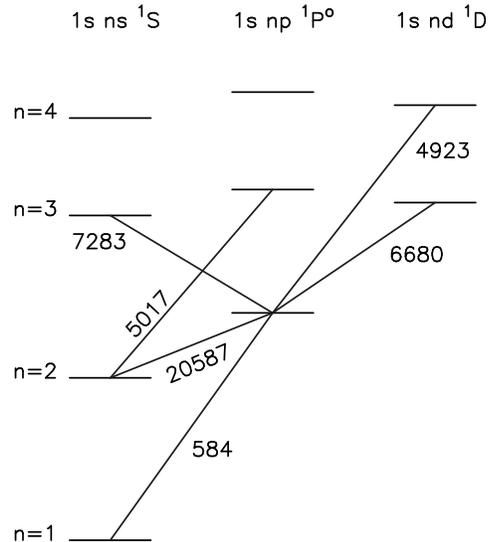}
\caption[]{Simplified Grotrian diagram for
the singlet states of \ion{He}{i}. The $\lambda 584$
transition plays a key role in determining the
population of the \singp\ level, and hence the strength of
transitions that involve that level. 
Although not shown, 
another important line often used
as a diagnostic ($\lambda$4389; \singp-1s 5d $^1$D)
also involves the \singp\ state.}
\label{Fig_grot}
\end{figure}

A simplified Grotrian diagram for the \ion{He}{i} atom (singlets
only) is shown in Fig.~\ref{Fig_grot}. For the present discussion the
most important transition is the 1s$^2$\,$^1$S-- \singp\ transition at
584\,\AA.  Key transitions, often used for diagnostic
purposes, are indicated.
Overlapping the 584\,\AA\ transition are several \ion{Fe}{iv}\ transitions.
The data for the relevant \ion{Fe}{iv} transitions used in CMFGEN, taken from 
Kurucz \& Bell (\citeyear{Kur95}), is provided in Table 1 ---
only the two most important transitions are shown.
Wavelengths for the \ion{Fe}{iv} transitions were calculated
using the known energy levels given in the NIST Atomic Spectra Database
(http://physics.nist.gov/PhysRefData/ASD/index.html), 
and are believed to be accurate.\footnote{
The energy values for the relevant lower levels in the NIST table
are quoted to 0.1\,cm$^{-1}$, while
the upper levels are quoted to an accuracy of 0.01\,cm$^{-1}$.
An accuracy of 0.1\,cm$^{-1}$ in the energy
difference between the two levels corresponds to less than 
0.2\,\kms, and thus it is unlikely that the wavelengths are significantly in error.} 
The velocity separation of the two
\ion{Fe}{iv}\ lines from the  \ion{HeI}\ $\lambda$584 is given, as is
the oscillator strengths computed by Becker \& Butler (\citeyear{BB95_FeIV}).
These show
significant differences to those of Kurucz \& Bell (\citeyear{Kur95}), 
and indicate that the oscillator strengths for these transitions must be regarded
as uncertain.

In O stars \ion{He}{i}\ 584\,\AA\ is often optically thick, so that photons
undergo a number of scatterings within the line before
escaping. Because the optical depth is large, a high population
of the \singp\ state is maintained enhancing the
strength of absorption lines ending on this state. However,
as the photons scatter other mechanisms come into play. In particular, they 
can be absorbed by the overlapping \ion{Fe}{iv}\ lines. As the upper states
of these levels have alternative decay routes, the absorbed photons
can be removed from the 584 transition, acting as a drain, and causing
the \singp \ population to be lowered. This leads to a
weakening of the transitions whose lower state is \singp. Clearly
the importance of this effect in model atmosphere calculations
will depend on whether the explicit interaction between the \ion{He}{i}\ 
resonance transition and the \ion{Fe}{iv}\ lines is taken into account, the accuracy
of the \ion{Fe}{iv}\ wavelengths, oscillator strengths and branching
ratios, and on other atmospheric parameters such as the
adopted microturbulent velocity. It is also obvious that the
importance of the effect will depend on the 
actual stellar parameters (\ie\ \Teff, $\log g$, $\Mdot$).

In this study we concentrate specifically on the influence of
the two \ion{Fe}{iv} lines on the \ion{He}{i}\
line strengths. It should be stressed, however, that other
\ion{Fe}{iv}\ lines, and lines due to other species 
(e.g., \ion{Ca}{v}, \ion{Fe}{vi}) could 
also potentially have an influence (albeit generally weaker) on
the \ion{He}{i}\ singlet line strengths. 

%In one model there were 13 metal lines (although some belonging to
%high ionization species) located
%within $\pm$ 30\kms\ of the \ion{He}{i}\ transition.

\begin{table*}
\caption{Atomic Data\label{tab1}}
\begin{tabular}{llllllll} \hline\hline
Ion &Low level$^1$ &  Upper level             & $\lambda$(\AA) & $\Delta V($km\,s$^{-1})^2$ & $A_{ul}$  &  $f_{lu}^3$  &  $f_{lu}^4$ \\
\hline
\ion{He}{i} &1s 2s $^1$S        &  1s 2p $^1$P$^{o}$ &  20586.9 & & $1.73 \times 10^6$ & 0.329  \\
\ion{He}{i} & 1s 1s $^1$S        &  1s 2p $^1$P$^{o}$ &  584.334  & & $1.78 \times 10^9$ & 0.274 \\
\ion{Fe}{iv} & 3d$^5$ $^4$F$_{9/2}$  &  3d$^4$($^3$G)4p $^2$H$^{o}_{9/2}$ & 584.368 & 17 & $6.82 \times 10^7$ & 0.00349 & 0.00251 \\
\ion{Fe}{iv} &3d$^5$ $^2$D3$_{5/2}$ &  3d$^4$($^3$G)4p $^4$H$^{o}_{7/2}$ & 584.397 & 32 & $4.22 \times 10^8$ & 0.0288  & 0.00264 \\
\hline
\end{tabular}
\hbox{ }\\
1. Level designations for \ion{Fe}{iv}\ are from the NIST
Atomic Spectra Database. The D3 designation is
used to denote one of the three 3d$^5$\,$^2$D terms. \\
2. Velocity shift of \ion{Fe}{iv}\ transition relative to \ion{He}{1}\ $\lambda 584$ transition.\\
3. \ion{Fe}{iv}\ oscillator strengths are from Kurucz \& Bell (\citeyear{Kur95}). \\ 
4. \ion{Fe}{iv}\ oscillator strengths are from Becker \& Butler (\citeyear{BB95_FeIV}).
\end{table*}

\section{Results}

In order to quantitatively explore the influence of the \ion{Fe}{iv} lines on
the \ion{He}{i} lines we consider the following model:
  $\Teff=33,560$\,K, 
  $\log g= 3.9$,
  $L =7.13 \times 10^4 \,\Lsun$,
  $\Rstar=7.8\,\Rsun $,
  $\Mdot=1.0 \times 10^{-8}\, \Msunyr$,
  $\Vinf=1400 \,\kms$.
The parameters of the model are similar to models we are
using to investigate the O9V star 10 Lac (Lanz \etal\ \citeyear{Lanz_10Lac}).
Since the mass-loss is low, the important parameters for the present study
are $\Teff$ and $\log g$ only. The hydrostatic structure  was initially
obtained from a TLUSTY model but has been subsequently modified
as the parameters were changed. Below 5\,\kms\ hydrostatic equilibrium
is achieved to generally better than 5\%. Unless otherwise stated, we
used a fixed Doppler width of 13.5\,\kms\ in the atmosphere calculations,
while a turbulent velocity of 10\,\kms\ was adopted for the profile
calculations.
The N(He)/N(H) ratio is taken as 0.1 while we adopted solar abundances
using the compilation of Grevesse \& Saval \cite{GS98_abund}. 
CNO abundances were reduced slightly in
accord with the recent results by Asplund \etal\ \cite{AGS06_abund}.

\subsection{\ion{He}{i} singlet sensitivities: $f$ values, abundances, and turbulent velocities}

\begin{figure}
\includegraphics[angle=270.0, scale=0.7]{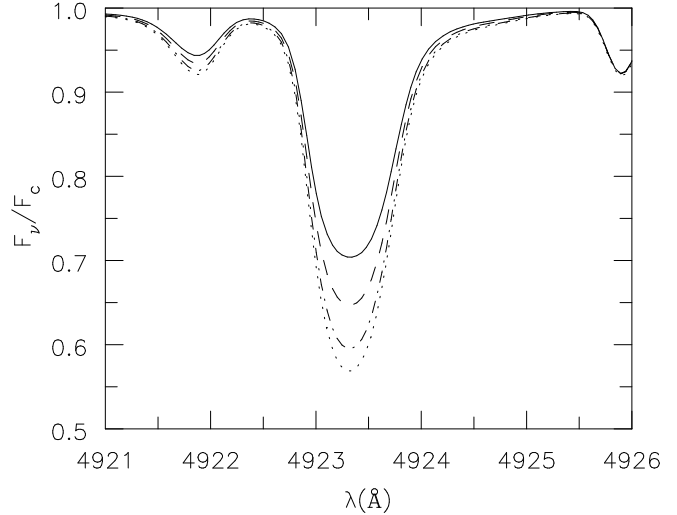}
\caption[]{Predicted line profiles for \ion{He}{i}\
$\lambda 4923$ (2p $^1$P$^o$-4d $^1$D). The solid curve show
the predictions with the standard model, while the
other curves show the profiles when 
the relevant \ion{Fe}{iv}\ $f$ values are reduced
by a factor of 2 (dashed curve), a factor of 5
(dot-dash), and a factor of 10 (dot).}
\label{Fig_4923}
\end{figure}

\begin{figure}
\includegraphics[angle=270.0, scale=0.7]{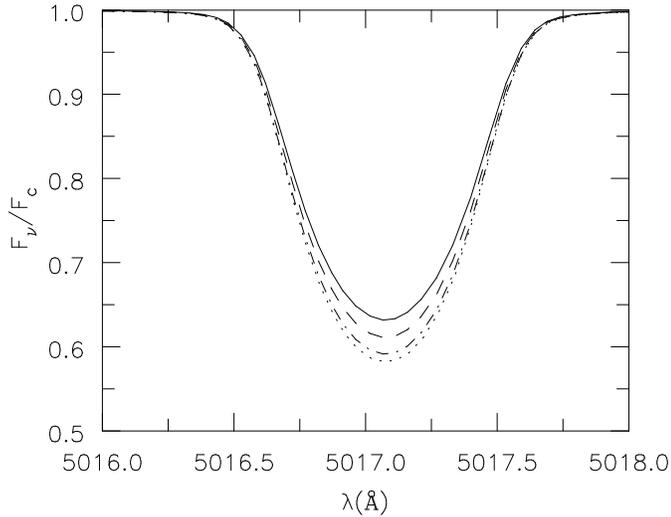}
\caption[]{As for Fig. \ref{Fig_4923} but for \ion{He}{i}\
$\lambda$5017 (1s 2s $^1$S-1s 3p $^1$P$^o$). The line
strengths have changed significantly, but the change is less
than for transitions ending on \singp.}
\label{Fig_5017}
\end{figure}

\begin{figure}
\includegraphics[angle=270.0, scale=0.7]{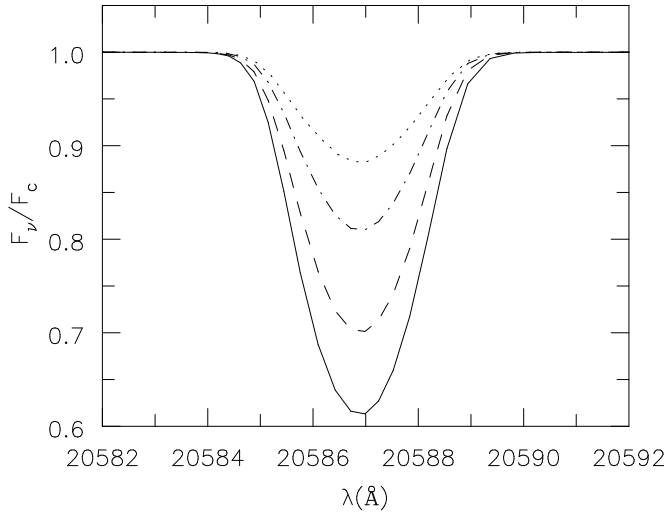}
\caption[]{As for Fig. \ref{Fig_4923} but for \ion{He}{i}\
$\lambda$20587 (2s $^1$S-2p $^1$P$^o$). Because the
\singp\ state is the upper level of this transition, the
variations in the strength are in the opposite sense to
the optical lines.}
\label{Fig_2058}
\end{figure}

\begin{figure}
\includegraphics[angle=270.0, scale=0.7]{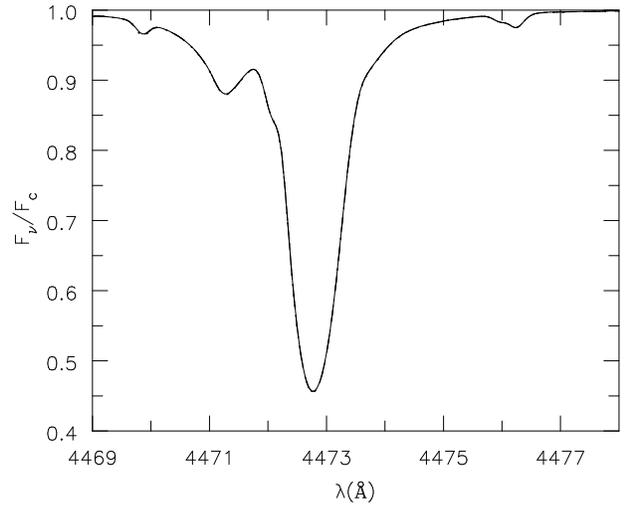}
\caption[]{As for Fig. \ref{Fig_4923} but for \ion{He}{i}\
$\lambda$4473 (2p $^3$P$^o$-4s $^3$D). The triplet
line strengths are not affected by changes in the
\ion{Fe}{iv}\ line strengths.}
\label{Fig_4473}
\end{figure}

In order to test the sensitivity of the models to the oscillator strengths
of the two relevant \ion{Fe}{iv}\ transitions we ran a detailed CMFGEN model
using the tabulated $f$ values, and then models in which the $f$ values 
were reduced by factors of 2, 5, and 10. The results for \ion{He}{i}\ 
$\lambda 4923$ (vacuum), $\lambda 5017$, $\lambda 20587$, and $\lambda 4473$, 
are shown in Figures \ref{Fig_4923}, \ref{Fig_5017},
\ref{Fig_2058} and \ref{Fig_4473} respectively.  
The three singlet lines show significant variations.
For the transition with \singp\ as the  lower level, the line is strongest 
in the model with the lowest $f$ values.
As noted earlier \ion{Fe}{iv}\ transitions drain photons from the
\gnds\ -- \singp\ transition, lowering the equilibrium \singp\ population.
When the oscillator strength is lowered the drain is weakened, and
the \singp\ population is enhanced. The $\lambda 4923$ transition,
which is coupled directly to the \singp\ state, shows a
large variation (Fig.~\ref{Fig_4923}), while the $\lambda 5017$ line
(Fig.~\ref{Fig_5017}), which has 2s $^1$S as its lower state, shows smaller variations.
Another transition, \ion{He}{i} $\lambda 6680$, which also
has \singp\ as its lower state, shows similar or even more
sensitive behavior than $\lambda 4923$. The near IR transition, 
\ion{He}{i}\ $\lambda 20587$ (1s\,2s\,$^1$S-\singp), also shows
very significant variations. However, in this case, the absorption 
profiles are strongest when the $f$ values are largest. This is to be
expected, since for this transition the \singp\ state is the upper
level. The difficulty in modeling the $\lambda 20587$ line in
early type stars has been discussed extensively, for example, by Najarro \etal\
(\citeyear{NHK94_GC}). The problem with $\lambda 20587$ has not been 
fully solved with models accounting for blanketing. In their analysis of 
WR147, Morris et al. \cite{MHC00_WR147} found that this line was one of the very few 
lines for which a discrepancy still appeared. They obtained a much too deep
\ion{He}{1}\ 20587 line, consistent with the singlet problem presented here.
The \ion{He}{1}\ $\lambda 20587$ line is very important for analyses of O stars
using H and K band infrared data (Repolust et al. \citeyear{RPH05_IR}), and hence
it is important that its formation be understood.

The triplet line, \ion{He}{i}\ $\lambda 4473$,
is essentially identical in the 4 models (Fig.~\ref{Fig_4473}). 
Similar statements also apply to other \ion{He}{i}\ 
triplet lines (e.g., $\lambda 5877$). Limited tests with TLUSTY and
FASTWIND confirm the changes found using CMFGEN.

\begin{figure}
\includegraphics[angle=270.0, scale=0.7]{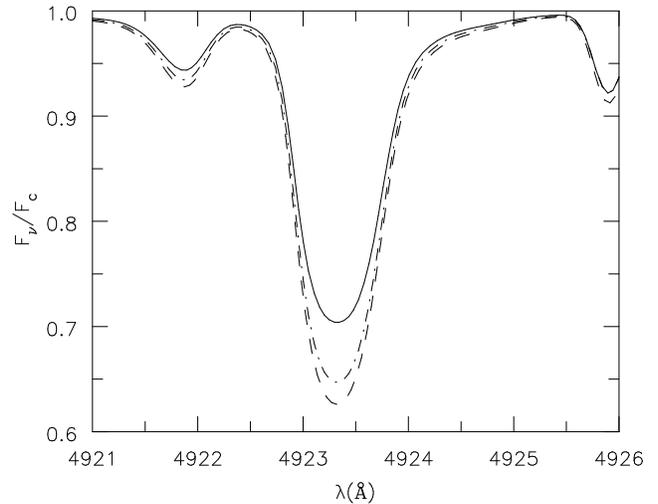}
\caption[]{Predicted line profiles for \ion{He}{i}\
$\lambda 4923$ (2p $^1$P$^o$-4d $^1$D). The solid curve show
the predictions with the standard model, the dashed curve
for a factor of 2 reduction in the Fe abundance, 
and the dot-dash curve for a factor of 2 reduction in
the relevant \ion{Fe}{iv}\ $f$ values.}
\label{Fig_abund_var}
\end{figure}

Another obvious way to reduce the ``drain" is to lower the 
Fe abundance. Indeed, as seen from Fig.~\ref{Fig_abund_var}, a
reduction in the Fe abundance by a factor of 2 gives a similar
enhancement in line strength in $\lambda 4923$ as does the
factor of 2 reduction in the Fe line strengths. The
correspondence is not exact since a lower Fe abundance reduces
line blanketing and hence causes slight changes in the temperature
structure of the model atmosphere. As before, the $\lambda 4473$
profiles are very similar in the three models. The sensitivity of the
singlet lines to the Fe abundance explains why Hillier \& Lanz
(\citeyear{HL01}) achieved excellent agreement when comparing 
models for SMC O stars where the Fe abundance is only 0.2 solar.

\begin{figure}
\includegraphics[angle=270.0, scale=0.7]{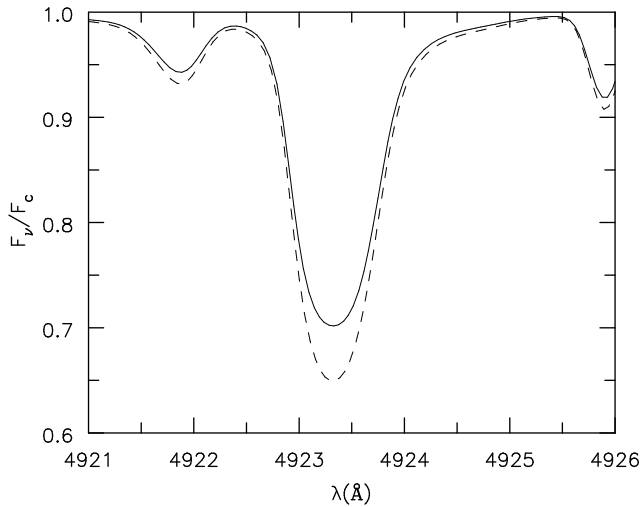}
\caption[]{Predicted line profiles for \ion{He}{i}
$\lambda 4923$ (2p $^1$P$^o$-4d $^1$D). The solid curve shows
the predictions with the standard model while the
dashed curve shows the effect of a lower turbulent
velocity in the calculation of the level populations.}
\label{Fig_turb}
\end{figure}

As a final test of the sensitivity of the \ion{He}{i}\ line
strengths we examined the effect of the microturbulent velocity used
in the atmosphere structure calculations to determine the
level populations. In Fig.~\ref{Fig_turb} we show the
profiles for $\lambda 4923$ computed using the
same turbulent velocity for the profile calculations
(10\,\kms) but computed using fixed Doppler widths of
10 and 13.5\,\kms\ for the determination of the atomic
populations. As might be expected, $\lambda 4923$ is
stronger when computed using the lower turbulent velocity.
With the larger turbulent velocity there is better overlap
between the \ion{Fe}{iv}\ and the \ion{He}{i}\ line, 
and hence a more effective drain.

The use of 10\,\kms\ for population calculations in O stars
is not unusual. Detailed analysis suggest that this
turbulent velocity is fairly typical, with possible
lower values for dwarfs, and somewhat higher values for 
supergiants. The precise physical cause/meaning of 
microturbulent velocities derived from profile fitting 
is unclear.

\begin{figure}
\includegraphics[angle=270.0, scale=0.7]{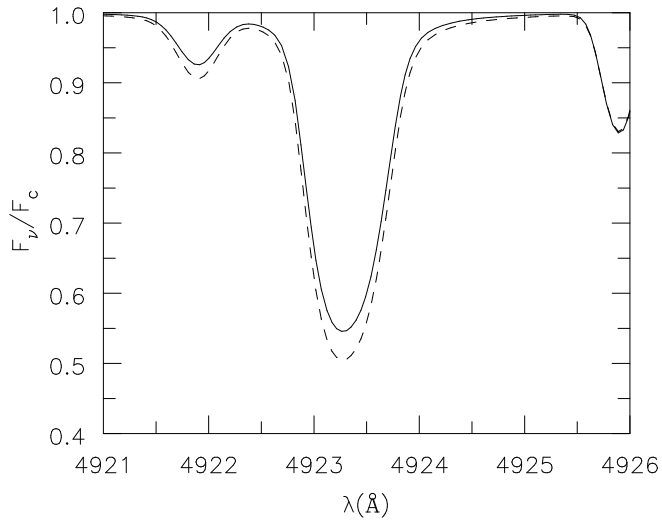}
\caption[]{Predicted line profiles for \ion{He}{i}\
$\lambda 4923$ (2p $^1$P$^o$-4d $^1$D) in a model for
a late O supergiant. The solid curve shows
the predictions with the standard model, while the
other curve shows the profile when
the relevant \ion{Fe}{iv}\ $f$ values are reduced
by a factor of 10.}
\label{Fig_losg}
\end{figure}

\begin{figure}
\includegraphics[angle=270.0, scale=0.7]{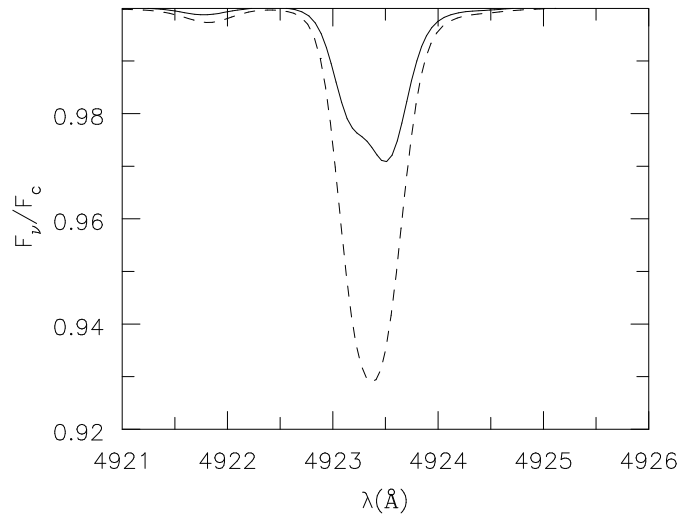}
\caption[]{Predicted line profiles for \ion{He}{i}\
$\lambda 4923$ (2p $^1$P$^o$-4d $^1$D) in a model for
an early O dwarf. The solid curve shows
the predictions with the standard model, while the
other curve shows the profile when
the relevant \ion{Fe}{iv}\ $f$ values are reduced
by a factor of 10.}
\label{Fig_eodw}
\end{figure}

We have also performed test calculations for other ``typical''
O star parameters. Similar effects are seen to occur,
although the size of the changes depends strongly on
spectral type. The mechanism does not appear to be
operative for B stars, and in a model of
a late O supergiant ($\Teff=28,400\,K, \log g=3.2$,
$\Mdot=1.65 \times 10^{-6}$\,\Msunyr, $L=5.4 \times 10^5$\,\Lsun)
the influence of the \ion{Fe}{iv}\ lines was rather small, as illustrated in 
Fig.~\ref{Fig_losg}. Conversely, the influence in an early O
dwarf ($\Teff=41,300$\,K, $\log g=4.0$, $\Mdot=2.0 \times 10^{-6}$
\,\Msunyr, $L=2.35 \times 10^5$\,\Lsun), where the \ion{He}{i}\ lines are 
weak, is relatively strong (Fig.~\ref{Fig_eodw}). 
Interestingly, in a mid-O supergiant
($\Teff=36,200$\,K,  $\log g=4.0$, $\Mdot=9.5 \times 10^{-7}$\,\Msunyr, 
$L=2.35 \times 10^5$\,\Lsun),
a reduction of a factor of 10 in the two \ion{Fe}{iv}\ 
oscillator strengths can cause \ion{He}{i}\ $\lambda 4923$
to change from being weakly in emission to a relatively strong 
absorption line.

In principle, $f$ values could be empirically
derived from a model comparison with observation.
However, given the problem with turbulent velocities,
and the sensitivities to the Fe abundance and other stellar 
parameters, we refrain from providing
actual values. However, it does appear that both the
Kurucz \& Bell (\citeyear{Kur95}) values, and the Becker \& Butler 
(\citeyear{BB95_FeIV}) values, are too large.

Taken as a whole, these findings explain the different 
results obtained between TLUSTY, CMFGEN, and FASTWIND.
While TLUSTY and CMFGEN use similar data for \ion{Fe}{iv}\
the models are not always computed using the same assumptions,
such as which species are included and the adopted turbulent 
velocity. In CMFGEN, a smaller turbulent velocity leads to a longer
computational time, and so a larger turbulent velocity is
often adopted. In general the choice has very little
effect of the population determinations but for a few lines,
as shown here, the adopted value can be important. The
adopted turbulent velocity was also shown to directly
affect the \ion{He}{ii}\ $\lambda 4687$ line in B supergiants
(Evans \etal\ \citeyear{ECF04_BSG}). In FASTWIND blanketing
is treated using mean opacities and emissivities, and 
consequently the detailed interaction between the
\gnds-\singp\ lines at 584\AA\ and the \ion{Fe}{iv}\
lines is not taken into account, although work is in progress to
rectify this. Since all three codes give similar answers
for the \ion{He}{i}\ triplet lines, the triplet
lines are clearly preferred for spectral analyses.

As noted by the referee, Herrero et~al.\ \cite{HPN02_OB}
were able to obtain a good fit to the \ion{He}{1} $\lambda 4473$,
and $\lambda  4389$ lines using a model with $\Teff=35,500 \pm 500\,$K 
and $\log g=3.95 \pm 0.1$. This can be explained by noting that in 10 Lac
it appears that the neglect of the influence
of the two \ion{Fe}{iv}\ transitions on the \ion{He}{i}\ singlet transitions is
a reasonable approximation. Finally, it should be noted, that
FASTWIND consistently gives temperatures, for O dwarfs, that are 
1500\,K higher than CMFGEN (and TLUSTY). The reason for the discrepancy 
is unknown but is being investigated.
%lines is not perfect (e.g., core of the \ion{He}{ii} $\lambda 4201$ is too
%weak in the model) suggesting the possibility of systematic errors,
%and that model improvements are still necessary. 
Mokiem et~al. \cite{MKP05_GA}
also obtained similar fit parameters to that of Herrero 
\etal ($\Teff=35,000 \pm 850\,$K
and $\log g=4.03 \pm 0.13$) using numerous FASTWIND models and
a sophisticated genetic algorithm. However, they derived a turbulent velocity
of $15.5 \pm 4\,$\kms\ which is inconsistent with
that indicated by the metal lines (Lanz \etal\ \citeyear{Lanz_10Lac}), 
again suggesting that possible systematic effects in the models
are being neglected (see
additional discussion in Lanz \etal\ \citeyear{Lanz_10Lac}).

\section{Conclusion}

Earlier work on the analysis of massive O stars 
revealed inconsistencies in the predictions of \ion{He}{i}\
line strengths with observations, and between different codes.
We have shown that these inconsistencies arise because
of the different treatment of line-blanketing in the neighborhood
of the \ion{He}{i} \gnds\ -- \singp\  transition at 584\,\AA,
due to different treatments regarding the interaction of the
\ion{He}{i}\ and \ion{Fe}{iv} model atoms, and
due to different adopted model \ion{Fe}{iv} atoms. 
The adopted microturbulent velocities and iron abundance
also play a direct role in determining the population of the
\singp\ state, and hence on the \ion{He}{i}\ singlet line strengths.

The accurate prediction of \ion{He}{i}\ lines, which involve
\singp, is difficult,
and possibly not feasible with the current atomic data
available for \ion{Fe}{iv}.  We argue strongly that the 
\ion{He}{i}\ triplet lines should be weighted the most heavily 
when performing spectral analyses.  Inconsistencies
between model fits of the \ion{He}{i}\ singlet lines and 
triplet lines, and between those singlets connected to
\singp (e.g., $\lambda 4923$) and other singlet lines
(e.g., $\lambda 5017$; 1s\,2s\,$^1$S-1s\,3p\,$^1$P$^o$)
are indicative of significant transfer effects in \ion{He}{i}\ $\lambda 584$. 

Extensive modeling by one of us (Najarro) suggests that the Kurucz \ion{Fe}{iv} 
oscillator strengths are too large but given the uncertainties
in the models, and the influence of the turbulent velocity,
it is not feasible to determine reliable empirical values.

%We have provided empirical data
%for two \ion{Fe}{iv}\ oscillator strengths which seem to give
%better consistency with observations, but caution against
%blind use. 

We have shown how subtle radiative
transfer effects can significantly influence the strength of
the \ion{He}{i}\ singlet transitions. It is possible that 
similar effects may occur for other diagnostic lines,
especially if the levels involved in such transitions
are coupled to low lying levels whose population may be effected
by overlap between strong transitions and metal lines.

Interestingly, the subtle effect of the \ion{Fe}{iv}\
lines on the \singp\ population may contribute to
the dilution effect. In early unblanketed O star models, 
triplet lines were consistently too weak compared with
observations, especially in stars of class II, Ib, and Ia
(Voels \etal\ \citeyear{VBA89_dil_eff}). This was interpreted as a problem with
the triplet lines, and was believed to arise from the
neglect of extension effects. Intriguingly, all the
singlet lines analyzed by Voels \etal\ \cite{VBA89_dil_eff}
involve the \singp\ state.
While it is now known that blanketing is much more important 
than was previously suspected, the use of spherical 
models, the allowance for stellar winds, and the
influence of line blanketing (as in FASTWIND) has not 
alleviated the problem with \ion{H}{1}\ $\lambda 4473$ in giants and
supergiants of type O6 and later (Massey \etal\ 
\citeyear{MPP05_LMC_O_stars}, Repolust \etal\ \citeyear{RPH04_Teff}).
Physically, and with the new generation of models, it is much 
easier to understand the inconsistencies in the \ion{He}{i}\ line strengths
as arising from problems in computing the population of
the \singp\ state and hence the singlet line strengths, 
rather than the triplet line strengths.

This work highlights the importance of performing analyses
with independent codes, and with performing systematic
analyses of all available lines. In this way discrepancies,
sensitivities, and biases can be determined and investigated.
The work also highlights the importance of understanding
the physics of line formation if one is to derive accurate
abundances. This is especially true in situations where only one
or two spectral lines are available for analysis --- a situation
not uncommon in O stars. 

\begin{acknowledgements}

The authors thank the referee,  Paul Crowther, for his
comments.
F.N. acknowledges PNAYA2003-02785-E and AYA2004-08271-C02-02 grants and 
the Ramon y Cajal program.
DJH gratefully acknowledges support from NASA LTSA grant
NAG5-8211 and NASA ADP grant NNG04GC81G.  
TL is supported by NASA ADP grant NNG04GC81G.
FM acknowledges support from the Alexander von Humboldt foundation. 

\end{acknowledgements}

%----------------------------------------------------------------------%
% Bibliography                                                         %
%----------------------------------------------------------------------%

\end{document}